# Lagrangian and Covariant Field Equations for Supersymmetric Yang-Mills Theory in 12D [1]

Hitoshi NISHINO

*Department of Physics*
*University of Maryland*
*College Park, MD 20742-4111, USA*
E-Mail: nishino@umdhep.umd.edu

## Abstract

We present a lagrangian formulation for recently-proposed supersymmetric Yang-Mills theory in twelve dimensions. The field content of our multiplet has an additional auxiliary vector field in the adjoint representation. The usual Yang-Mills field strength is modified by a Chern-Simons form containing this auxiliary vector field. This formulation needs no constraint imposed on the component field from outside, and a constraint on the Yang-Mills field is generated as the field equation of the auxiliary vector field. The invariance check of the action is also performed without any reference to constraints by hand. Even though the total lagrangian takes a simple form, it has several highly non-trivial extra symmetries. We couple this twelve-dimensional supersymmetric Yang-Mills background to Green-Schwarz superstring, and confirm fermionic $\kappa$-invariance. As another improvement of this theory, we present a set of fully Lorentz-covariant and supercovariant field equations with no use of null-vectors. This system has an additional scalar field, whose gradient plays a role of the null-vector. This system exhibits spontaneous breaking of the original Lorentz symmetry $SO(10,2)$ for twelve-dimensions down to $SO(9,1)$ for ten-dimensions.

---

[1] This work is supported in part by NSF grant # PHY-93-41926.

# 1. Introduction

The investigation of possible supersymmetric theories in twelve-dimensions (12D) with the signature $(-, +, \cdots, +, -)^2$ beyond 10D is strongly motivated by the recent development in F-theory [1][2][3], S-theory [4], or theories with two times [5]. These higher-dimensional theories in turn are motivated by M-theory [6] in 11D, namely by the indication that the perturbative and non-perturbative states of M-theory may be unified by possible superalgebra in $D = (10, 2)$ [7] or $D = (11, 3)$ [8]. Considering these recent developments in M-theory, F-theory, S-theory and theories in two times, it is imperative to establish first the explicit models with supersymmetry in these higher dimensions, that have definite field theoretic representations realizing expected algebras.

Based on this philosophy, we have presented explicit formulations of $N = 1$ supersymmetric Yang-Mills (SYM) theory [9], an $N = 1$ supergravity theory [10], an $N = 2$ supergravity theory [11] in 12D, and also $N = 1$ SYM theories in arbitrary even dimensions beyond 12D [12] as a generalization of the work in 11+3 dimensions [13]. In particular, the $N = 1$ supersymmetric formulation plays an important key role for understanding the basic structure of supersymmetry in 12D. The novel feature is a completely new supersymmetry algebra in 12D

$$\{Q_\alpha, Q_\beta\} = (\gamma^{\mu\nu})_{\alpha\beta} P_\mu n_\nu \tag{1.1}$$

involving a null-vector in an unconventional way [4][9]. This algebra may well be further generalized by replacing null-vectors by extra momenta [7][5][14] in higher dimensions [12]. Furthermore, since $N = 1$ SYM theory in 10D has no origin in 11D supergravity or M-theory [6], our newly developed $N = 1$ SYM theory in 12D is to be the fundamental theory for the 10D SYM theory. From this viewpoint, it is also important to study the consistency of SYM theory in 12D with 10D superstring theory.

Despite of all of these progresses, we still have some unsolved problems related to the supersymmetric theories in 12D or beyond [9][10][11][12]. For example, in ref. [9], supersymmetry is realized only by Lorentz non-covariant set of field equations, and we have not yet discovered any invariant lagrangian formulation for any of these theories in 12D. The recent development in M-theory [6] relating itself to matrix theory [15] strongly suggests the existence of invariant lagrangian, if 12D theories are more fundamental than 11D or 10D theories. A separate problem to be solved is the lack of covariant set of field equations, even in the absence of an invariant action.

In this paper we solve these problems of the lack of lagrangian and covariant field equations in 12D SYM theory, by presenting an explicit invariant lagrangian under supersymmetry, and a set of Lorentz covariant field equations in component formulation. For the lagrangian formulation, the invariance of the action under supersymmetry is confirmed with no use of constraints on fundamental field. We introduce a new auxiliary vector field that gauges the extra symmetry called $\Omega$-symmetry in [9]. This auxiliary vector field is introduced into the field strength of the usual Yang-Mills field as Chern-Simons form. As for the above-mentioned consistency between 12D SYM background and superstring, we confirm the fermionic $\kappa$-invariance in Green-Schwarz superstring formulation. We next present a covariant set of field equations at the expense of an invariant lagrangian. We use an additional scalar field that is intact under supersymmetry, whose gradient replaces the null-vectors everywhere in the formulation. We clarify how this formulation avoids the usual problem with the closure of supersymmetry, when a new scalar field is included.

---

[2]This 12D space-time is sometimes denoted by $D = (10, 2)$.

## 2. Lagrangian and Invariances

We first summarize our results in a compact way, and associated remarks will be given later. Our field content is similar to that already given in refs. [9][12], with only one additional auxiliary vector field $C_\mu{}^I$ in the adjoint representation, namely with the field content $(A_\mu{}^I, \lambda^I, C_\mu{}^I)$, where we use the indices $I, J, \cdots$ for the adjoint representation as in [9]. Other conventions such as the definition of $\gamma_{13}$ are the same as in [9], or self-explanatory. Our invariant lagrangian $\mathcal{L}_0$ under supersymmetry has only two explicit terms:

$$I_0 = \int d^{12}x\, \mathcal{L}_0 = \int d^{12}x\, \left[ -\tfrac{1}{4}(\widetilde{F}_{\mu\nu}{}^I)^2 + (\overline{\lambda}^I \gamma^{\mu\nu} D_\mu \lambda^I) n_\nu \right] \quad . \tag{2.1}$$

The covariant derivative $D_\mu$ has the usual minimal coupling to the gauge field: $D_\mu \lambda^I \equiv \partial_\mu \lambda^I + f^{IJK} A_\mu{}^J \lambda^K$ with the gauge group structure constants $f^{IJK}$, and this is common to any other (combinations of) fields carrying the adjoint indices. Here the 'modified' field strength $\widetilde{F}_{\mu\nu}{}^I$ has a Chern-Simons form involving the $C_\mu{}^I$-field:

$$\widetilde{F}_{\mu\nu}{}^I \equiv \partial_\mu A_\nu{}^I - \partial_\nu A_\mu{}^I + f^{IJK} A_\mu{}^J A_\nu{}^K - C_\mu{}^I n_\nu + C_\nu{}^I n_\mu \quad , \tag{2.2}$$

satisfying the Bianchi identity

$$D_{[\mu} \widetilde{F}_{\nu\rho]}{}^I \equiv -H_{[\mu\nu}{}^I n_{\rho]} \quad , \tag{2.3}$$

with the field strength $H_{\mu\nu}{}^I \equiv D_\mu C_\nu{}^I - D_\nu C_\mu{}^I$ of $C_\mu{}^I$ with the minimal coupling to $A_\mu{}^I$. Even with the null-vector involved, we regard the $C_\mu$-linear terms in the modified field strength as a Chern-Simons form, due to the appearance of the field strength $H_{\mu\nu}{}^I$ in the Bianchi identity. This is also related to the extra symmetry of the $A_\mu$-field compensated by the transformation of the $C_\mu$-field, as will be clarified shortly. Note that the $C_\mu$-field is involved only in the modified field strength, and is not explicit in any other term.

The supersymmetry transformation rule of our multiplet is[3]

---
[3]Even though this multiplet seems different from that in [9], there is a closer link between them. See the paragraph with eq. (2.16).

$$\delta_Q A_\mu{}^I = +(\overline{\epsilon}\gamma_\mu \slashed{n} \lambda^I) \ , \tag{2.4a}$$

$$\delta_Q \lambda^I = +\tfrac{1}{4}\gamma^{\mu\nu}\epsilon \widetilde{F}_{\mu\nu}{}^I \ , \tag{2.4b}$$

$$\delta_Q C_\mu{}^I = -(\overline{\epsilon}\gamma_\mu{}^\nu D_\nu \lambda^I) \ . \tag{2.4c}$$

As usual $n_\mu$ and $m_\mu$ are null-vectors with non-zero components only for the extra coordinates, satisfying $n^\mu n_\mu = 0$, $m^\mu m_\mu = 0$, $n^\mu m_\mu = +1$, $n_+ = m^+ = +1$, $n_- = m^- = 0$ [9][10], and $\slashed{n} \equiv \gamma^\mu n_\mu$, $\slashed{m} \equiv \gamma^\mu m_\mu$. The supersymmetric invariance of our action (2.1) is easily confirmed, by the aid of the l.h.s. of the following field equations,[4] obtained respectively by varying the action with respect to $A_\mu$, $\lambda$ and $C_\mu$:

$$D_\nu \widetilde{F}^{\mu\nu\,I} + f^{IJK}(\overline{\lambda}^J \gamma^{\mu\nu}\lambda^K)n_\nu = 0 \ , \tag{2.5}$$

$$\gamma^{\mu\nu}D_\mu \lambda^I n_\nu = 0 \ , \tag{2.6}$$

$$\widetilde{F}_{\mu\nu}{}^I n^\nu = 0 \ . \tag{2.7}$$

Note that the leading term in (2.5) has no longer the null-vector in this formulation, unlike that in ref. [9].

The closure of supersymmetry is highly non-trivial. Our result of closure is summarized as

$$[\,\delta_Q(\epsilon_1), \delta_Q(\epsilon_2)\,] = \delta_P + \delta_\Lambda + \delta_\Omega + \delta_\Sigma + \delta_\zeta + \delta_\eta \ , \tag{2.8}$$

where $\delta_P$ is the usual translation with the parameter $\xi^\mu \equiv (\overline{\epsilon}_1 \gamma^{\mu\nu}\epsilon_2)n_\nu$ [9], and $\delta_\Lambda$ is the usual Yang-Mills gauge transformation with the parameter $\Lambda^I$, while $\delta_\Omega$, $\delta_\Sigma$, $\delta_\zeta$, $\delta_\eta$ are extra symmetries that are inherent in our system, similarly to refs. [9][12]. They are highly non-trivial symmetries dictated by

$$\delta_\Omega A_\mu{}^I = n_\mu \Omega^I \ , \quad \delta_\Omega \widetilde{F}_{\mu\nu}{}^I = 0 \ , \tag{2.9a}$$

$$\delta_\Omega C_\mu{}^I = D_\mu \Omega^I \ , \tag{2.9b}$$

$$\delta_\Sigma C_\mu{}^I = n_\mu \Sigma^I \ , \quad \delta_\Sigma \widetilde{F}_{\mu\nu}{}^I = 0 \ , \tag{2.10}$$

$$\delta_\zeta C_\mu{}^I = -\tfrac{7}{16} f^{IJK}(\zeta^{\rho\nu}\lambda_{\rho\mu}{}^{JK}n_\nu + \zeta_{\mu\nu}\lambda^{\nu\rho\,JK}n_\rho) \ ,$$
$$\delta_\zeta \lambda^I = -\tfrac{7}{8}\zeta^{\mu\nu}(\gamma_{\rho\mu}D_\nu\lambda^I n^\rho + \gamma_\mu{}^\rho D_\rho \lambda^I n_\nu) \ , \tag{2.11}$$

$$\delta_\eta C_\mu{}^I = \tfrac{3}{16(6!)} f^{IJK}(\eta^{[5]}{}_\mu \lambda_{[5]\rho}{}^{JK}n^\rho - \eta^{[5]\rho}\lambda_{[5]\mu}{}^{JK}n_\rho) \ ,$$
$$\delta_\eta \lambda^I = -\tfrac{1}{64(6!)}\eta^{[6]}\gamma^{\mu\nu}\gamma_{[6]}D_\mu \lambda^I n_\nu \ , \tag{2.12}$$

---

[4]To avoid misunderstanding, note that we use only the 'l.h.s.' of these equations for taking variations, but *not* the field equations themselves! If the field equations were used in invariance check, any lagrangian would be trivially invariant, as is well-known in field theory.

where the infinitesimal arbitrary parameters $\Omega^I$ and $\Sigma^I$ are local, while $\zeta^{\mu\nu}$ and $\eta^{[6]}$ are global.[5] We use the universal convention $\lambda_{\mu\nu}{}^{IJ} \equiv (\bar{\lambda}^I \gamma_{\mu\nu} \lambda^J)$, $\lambda_{[6]}{}^{IJ} \equiv (\bar{\lambda}^I \gamma_{[6]} \lambda^J)$ in this paper. Also used is the shorthand notation $[n]$, denoting $n$-th rank totally antisymmetric indices. Eq. (2.9b) implies that $C_\mu{}^I$ is the gauge field for the $\Omega$-symmetry. In the commutator algebra (2.8) these parameters are found to be

$$\Lambda^I = -\xi^\mu A_\mu{}^I \ , \tag{2.13a}$$

$$\Omega^I = -\tfrac{1}{2}\zeta^{\mu\nu} \widetilde{F}_{\mu\nu}{}^I \ , \tag{2.13b}$$

$$\Sigma^I = -\tfrac{1}{2}\zeta^{\rho\sigma} H_{\rho\sigma}{}^I + \tfrac{7}{32} f^{IJK}\zeta^{\rho\sigma}(\bar{\lambda}^J \gamma_{\rho\sigma} \lambda^K) - \tfrac{3}{8(6!)} f^{IJK} \eta^{[6]} \lambda_{[6]}{}^{JK} \ , \tag{2.13c}$$

$$\zeta^{\mu\nu} \equiv (\bar{\epsilon}_1 \gamma^{\mu\nu} \epsilon_2) \ , \quad \eta^{[6]} \equiv (\bar{\epsilon}_1 \gamma^{[6]} \epsilon_2) \ . \tag{2.13d}$$

The closure confirmation is straightforward, once we understand the result (2.8) - (2.13) showing how these extra symmetries are involved. Another crucial non-trivial confirmation is the invariance of our lagrangian (action) of (2.1) under the extra symmetries, in particular $\delta_\zeta$ and $\delta_\eta$. Here we give some crucial relationships we used in the confirmations. In the $\delta_\zeta$-invariance we use the crucial identity:

$$(D_\mu \bar{\lambda}^I) \gamma^{\rho\sigma} (D_\nu \lambda^I) = -\tfrac{1}{2} f^{IJK} (\bar{\lambda}^I \gamma^{\rho\sigma} \lambda^J) F_{\mu\nu}{}^K \ , \tag{2.14}$$

up to a total divergence. In the $\delta_\eta$-invariance, we use the important feature that any product of self-dual and anti-self-dual six-th rank tensors in 12D vanishes:[6]

$$S_{[6]} A^{[6]} \equiv 0 \ , \tag{2.15}$$

where $S_{[6]} \equiv +(1/6!)\epsilon_{[6]}{}^{[6]'} S_{[6]'}$ and $A_{[6]} \equiv -(1/6!)\epsilon_{[6]}{}^{[6]'} A_{[6]'}$. Using this, we see for example that $\eta^{[6]}(\bar{\lambda}^J \gamma^\rho \gamma_{[6]} \gamma^\sigma \lambda^K) \equiv 0$, because $\eta^{[6]}$ is anti-self-dual, while the remainder is self-dual with respect to the indices $[6]$.

Note that our lagrangian is actually invariant under these highly non-trivial extra symmetries, as well as supersymmetry, despite of the simplicity of its structure. This feature of extra symmetries is in a sense similar to those in Chern-Simons theories in 3D [17], where many such extra symmetries of the lagrangian show up in the commutator algebra of supersymmetries, which are sometimes implicit and easily overlooked in ordinary non-supersymmetric theories.

There are other remarks associated with our result. We first elucidate the relationship of the present multiplet with the original one in [9][12], which has not been clear so far. In refs. [9][12], the supersymmetry transformation rule for a vector multiplet $(A_\mu{}^I, \lambda^I)$ in

---

[5]The particular coefficients, such as $3/(16 \cdot 6!)$ is for normalizations complying with the closure of algebra (2.8).

[6]This situation is opposite to, *e.g.*, 6D case [16], where we have $S_{[3]} S^{[3]} \equiv 0$, $A_{[3]} A^{[3]} \equiv 0$, instead.

$D = (10, 2)$ [9] is given by

$$\delta_Q A_\mu{}^I = +(\overline{\epsilon}\gamma_\mu \lambda^I) \ ,$$
$$\delta_Q \lambda^I = +\tfrac{1}{4}\gamma^{\mu\nu\rho}\epsilon F_{\mu\nu}{}^I n_\rho = +\tfrac{1}{4}\slashed{n}\left(\gamma^{\mu\nu} F_{\mu\nu}{}^I\right) \ , \quad (2.16)$$

where the null-vector appears in the gaugino transformation rule. As shrewd readers may have already noticed, we can rewrite the second line as $\delta_Q \lambda^I = \tfrac{1}{4}\slashed{n}\gamma^{\mu\nu} F_{\mu\nu}{}^I$, under our extra constraint $F_{\mu\nu}{}^I n^\nu = 0$ [9]. This suggests an an alternative expression of this multiplet using the new field $\chi$ such that $\lambda \equiv \slashed{n}\chi$. In other words, we have an alternative SYM multiplet

$$\delta_Q A_\mu{}^I = +(\overline{\epsilon}\gamma_\mu \slashed{n} \lambda^I) \ , \quad (2.17\text{a})$$
$$\delta_Q \lambda^I = +\tfrac{1}{4}\gamma^{\mu\nu}\epsilon \, F_{\mu\nu}{}^I \ . \quad (2.17\text{b})$$

Here we used the same symbol $\lambda$ for the gaugino field instead of $\chi$, once the above-mentioned replacement of $\lambda$ by $\chi$ was made. Note here that the new gaugino field in (2.17) is unconstrained and it needs no extra constraint such as $\slashed{n}\lambda^I = 0$, as opposed to the case of (2.16) [9]. This transformation rule (2.16) is prototype of (2.4a) and (2.4b), before the auxiliary field $C_\mu$ is introduced.

We mention an interesting fact that the kinetic operator for the gaugino in our lagrangian (2.1) is nothing else than a 'generalized' Dirac operator in exactly the same form as the r.h.s. of our supersymmetry algebra (1.1). This indicates that the present formulation is more natural than the previous one in [9], due to the generalized Dirac operator in the gaugino kinetic term.

We next study the physical degrees of freedom for our gaugino field which seem non-trivial. This is because our $\lambda$ has no such a constraint as $\slashed{n}\lambda = 0$ as opposed to that in ref. [9]. However, we can understand that the actual physical degrees of freedom of $\lambda$ is reduced into the half of its original value as a Majorana-Weyl spinor, by considering the following extra symmetry:

$$\delta_\alpha \lambda^I = P_\uparrow \alpha^I \ , \quad (n^\mu D_\mu \alpha^I = 0) \ . \quad (2.18)$$

which can be easily shown to leave our action (2.1) invariant. The condition in the parentheses is analogous to eq. (12) in ref. [9]. The significance of this symmetry is rather transparent. First, notice that $P_\uparrow \equiv 2^{-1}\slashed{n}\slashed{n}$ is the same projection operator defined in [10]. Therefore eq. (2.18) implies that half of the original degrees of freedom of $\lambda$ in the direction of $P_\downarrow \equiv 2^{-1}\slashed{n}\slashed{n}$ are 'gauge' degrees of freedom, which are definitely non-physical. Hence the original $2^{12/2-1-1} = 16$ on-shell degrees of freedom as a Majorana-Weyl spinor are reduced to be at most 8, in agreement with the conventional 10D SYM theory. As for the field $C_\mu$, its role is clear as an auxiliary field due to the lack of its kinetic term, while the field equation (2.7) deletes one of the extra components in the field strength: $\widetilde{F}_{\mu-} = 0$ in the notation in [10]. Note also that this system is *not* quite reduced to 10D SYM theory, because of the remaining extra components such as $\widetilde{F}_{\mu+}$ which are still non-vanishing. In this sense, our system is different from a rewriting of 10D SYM, or the latter in 'disguise'.

We mention one important feature related to our $C_\mu$-field. Note that the variation of $C_\mu n^\mu$ under supersymmetry vanishes on-shell by the use of $\lambda$-field equation, as seen from (2.6) and (2.4c). Due to this feature, we can add the action

$$I' \equiv \int d^{12}x\, \mathcal{L}' \equiv \int d^{12}x \left[ \tfrac{1}{2} L (C_\mu{}^I n^\mu)^2 \right] \tag{2.19}$$

to $I_0$, with a new auxiliary field $L$ intact under supersymmetry, while adding a new term proportional to this $L$ in $\delta_Q \lambda$:

$$\begin{aligned}
\delta_Q L &= 0 ~, \\
\delta_Q \lambda^I &= +\tfrac{1}{2} \gamma^{\mu\nu} \epsilon \widetilde{F}_{\mu\nu}{}^I - \tfrac{1}{2} \epsilon L (C_\mu{}^I n^\mu) ~.
\end{aligned} \tag{2.20}$$

The invariance of our total action $I_0 + I'$ under supersymmetry can be easily confirmed: The variation of $\mathcal{L}'$ under supersymmetry is only from $\delta_Q (C_\mu{}^I n^\mu)^2$ proportional both to the l.h.s. of $\lambda$-field equation and $C_\mu n^\mu$, which in turn is cancelled by the above additional term in $\delta_Q \lambda$. Also to be mentioned is invariances under other extra symmetries. First, $I'$ is invariant under $\delta_\Omega$, when the parameter $\Omega$ satisfies $n^\mu D_\mu \Omega^I = 0$. The invariance of $I'$ under $\delta_\Sigma$ is trivial, while the invariances under $\delta_\zeta$ and $\delta_\eta$ are less trivial, but straightforward. Finally, the $L$-field equation yields the new field equation

$$C_\mu{}^I n^\mu = 0 ~, \tag{2.21}$$

as long as the Yang-Mills gauge group is compact, while the field equations (2.5) - (2.7) for the previous set of fields are intact, because the only possible new contributions from $\mathcal{L}'$ are vanishing due to (2.21). Since the $L$-field is non-physical and completely decouples from all the field equations, the invariance of $L$ under supersymmetry does not pose any problem.

## 3. Green-Schwarz Superstring on SYM

Once the SYM theory in 12D is established, next natural question is whether it can couple consistency to superstring. In this paper we confirm such consistency by studying the fermionic $\kappa$-invariance for the action for Green-Schwarz superstring on the 12D SYM background, but with no supergravity background, i.e., on flat superspace.

We start with the arrangement of superspace constraints necessary for our fermionic $\kappa$-invariance of our Green-Schwarz superstring action. The only relevant ones are of dimensionalities $d \leq 1/2$, which are

$$T_{\alpha\beta}{}^c = (\gamma^{cd})_{\alpha\beta} n_d + (P_\uparrow - P_\downarrow)_{\alpha\beta} n^c = G_{\alpha\beta}{}^c ~, \tag{3.1a}$$

$$F_{ab}{}^{rs} = -(\gamma_b \gamma^c)_\alpha{}^\beta \lambda_\beta{}^{rs} n_c ~. \tag{3.1b}$$

We do not take into account the constrains associated with $N=1$ supergravity such as $T_{\alpha\beta}{}^\gamma$, because we deal only with flat superspace backgrounds. Since the Green-Schwarz formulation is based on target superspace, we use the index convention for the local Lorentz indices in superspace $A, B, \cdots$, such as $a, b, \cdots$ denote the local bosonic coordinates, while $\alpha, \beta, \cdots$ for local undotted fermionic coordinates in the 12D superspace. The indices $r, s, \cdots = 1, 2, \cdots, 32$ are for the vectorial representation **32** for the gauge group $SO(32)$ chosen for anomaly cancellation. Accordingly we use the same indices for the field strength superfield $F_{AB}{}^{rs}$ instead of $F_{AB}{}^I$ in this section. Eq. (3.1a) is the same as in [10], while (3.1b) is derived from component transformation (2.17a) by the universal technique in [18].

We next summarize our result for the total action $S$, which consists of four parts, namely in addition to the three parts $S_\sigma$, $S_B$ and $S_\Lambda$ in [10], we have an action $S_\Psi$ for unidexterous Majorana-Weyl fermions $\Psi^r$, similarly to 10D case [19]:

$$S \equiv S_\sigma + S_B + S_\Lambda + S_\Psi \ , \tag{3.2}$$

$$S_\sigma \equiv \int d^2\sigma \left[ V^{-1} \eta_{ab} \Pi_{\mathop{+}\limits^{\_}}{}^a \Pi_={}^b \right] \ , \tag{3.3}$$

$$S_B \equiv \int d^2\sigma \left[ V^{-1} \Pi_{\mathop{+}\limits^{\_}}{}^A \Pi_={}^B B_{BA} \right] \ , \tag{3.4}$$

$$S_\Lambda \equiv \int d^2\sigma \left[ V^{-1} \Lambda_{++}(\Pi_={}^a n_a)(\Pi_={}^b m_b) + V^{-1} \widetilde\Lambda_{++} \left\{ (\Pi_={}^a n_a)^2 + (\Pi_={}^a m_a)^2 \right\} \right] \ , \tag{3.5}$$

$$S_\Psi \equiv \int d^2\sigma \left[ V^{-1} \Psi_+^r V_={}^i \left( \partial_i \Psi_+^r + \Pi_i{}^B A_B{}^{rs} \Psi_+^s \right) \right] \ . \tag{3.6}$$

Here $\det(V_{(i)}{}^j)$ is the determinant of the 2D zweibein $V_{(i)}{}^i$, where the indices $i, j, \cdots = 0, 1$ are for the curved 2D coordinates $\sigma^i$, while $(i), (j), \cdots = +, =$ are for the local Lorentz frames in 2D. The reason we need the 'doubled' $+$ or $=$ signs is to comply with the Lorentz charge for our fermion $\Psi_+^r$ carrying only one $+$. As is always the case in 2D, the sum of $+$'s and $-$'s within a lagrangian is supposed to cancel, as an equivalent statement to Lorentz invariance. Other notations are usual, such as $\Pi_i{}^A \equiv (\partial_i Z^M) E_M{}^A$ with 12D superspace coordinates $Z^M$ and inverse vielbein $E_M{}^A$ used for 'pull-backs'.

We can show the invariance of this total action under the following fermionic $\kappa$-symmetry, in a way similar to the proof given in [10]:

$$\delta_\kappa V_+{}^i = \overline\kappa_{\mathop{+}\limits^{\_}}{}^{\dot\alpha} (\gamma^c)_{\dot\alpha}{}^\beta \Pi_{\mathop{+}\limits^{\_}\beta} n_c V_={}^i - \tfrac{1}{2} \overline\kappa_{\mathop{+}\limits^{\_}}{}^{\dot\alpha} (\gamma^c)_{\dot\alpha}{}^\beta \lambda_\beta{}^{rs}(\Psi_+^r \Psi_+^s) n_c$$

$$\equiv (\overline\kappa_{\mathop{+}\limits^{\_}} \not{n} \Pi_{\mathop{+}\limits^{\_}}) V_={}^i - \tfrac{1}{2} (\overline\kappa_{\mathop{+}\limits^{\_}} \not{n} \lambda^{rs})(\Psi_+^r \Psi_+^s) \ , \quad (\not{n})_\alpha{}^{\dot\beta} \overline\kappa_{\mathop{+}\limits^{\_}\dot\beta} \equiv (\not{n} \overline\kappa_{\mathop{+}\limits^{\_}})_\alpha = 0 \ , \tag{3.7a}$$

$$\delta_\kappa V_={}^i = 0 \ , \quad \delta_\kappa(V^{-1}) = 0 \ , \quad \delta_\kappa \overline E^{\dot\alpha} = \delta_\kappa E^a = 0 \ , \tag{3.7b}$$

$$\delta_\kappa E^\alpha = \tfrac{1}{2} (\gamma_a)^{\alpha\dot\beta} \overline\kappa_{\mathop{+}\limits^{\_}\dot\beta} \Pi_={}^a \equiv \tfrac{1}{2} (\not{\Pi}_= \overline\kappa_{\mathop{+}\limits^{\_}})^\alpha \ , \tag{3.7c}$$

$$\delta_\kappa \Lambda_{++} = -2 (\overline\kappa_{\mathop{+}\limits^{\_}} \not{n} \Pi_{\mathop{+}\limits^{\_}}) \ , \quad \delta_\kappa \widetilde\Lambda_{++} = 0 \ , \tag{3.7d}$$

$$\delta_\kappa \Psi_+^r = -(\delta_\kappa E^\alpha) A_\alpha{}^{rs} \Psi_+^s \ , \tag{3.7e}$$

$$\delta_\kappa A_B{}^{rs} = (\delta_\kappa E^\gamma) E_\gamma{}^M \partial_M A_B{}^{rs} \ , \tag{3.7f}$$

where we use essentially the same notation as in [10], except those terms with new fermions $\Psi^r$.

We now confirm the $\kappa$-invariance of the total action. First of all, we do not repeat those cancellations described in [10] before adding our fermions $\Psi^r$. The only important new contributions are from the variation of $S_\Psi$ and the new $\Psi$-dependent term in $\delta_\kappa V_{\mp}{}^i$. As in the case of 10D Green-Schwarz superstring [19], we can see that these two contributions cancel each other. To be more specific, the former is arranged as

$$\begin{aligned}\delta_\kappa S_\Psi &= +(\delta_\kappa E^C)\Pi_{\mp}{}^B F_{BC}{}^{rs}(\Psi^r_+\Psi^s_+) \\ &= -\tfrac{1}{2}(\slashed{\Pi}_=\overline{\kappa}_{\mp})^\gamma \Pi_={}^b (\gamma_b\gamma^c\lambda^{rs})_\gamma (\Psi^r_+\Psi^s_+)n_c \\ &= -\tfrac{1}{2}(\overline{\kappa}_{\mp}\slashed{n}\lambda^{rs})\Pi_={}^a\Pi_{=a}(\Psi^r_+\Psi^s_+) \quad ,\end{aligned} \quad (3.8)$$

while the latter is from the new term in $\delta_\kappa V_{\mp}{}^i$ in $\delta_\kappa S_\sigma$:

$$\begin{aligned}\delta_\kappa S_\sigma|_{\text{new}} &= +(\delta_\kappa V_{\mp}{}^i)|_{\text{new}} V_i{}^{\mp}\Pi_={}^a\Pi_{=a} \\ &= +\tfrac{1}{2}(\overline{\kappa}_{\mp}\slashed{n}\lambda^{rs})\Pi_={}^a\Pi_{=a}(\Psi^r_+\Psi^s_+) \quad .\end{aligned} \quad (3.9)$$

Thus we can get the cancellation $\delta_\kappa(S_\sigma+S_\Psi) = 0$, and therefore the total invariance $\delta_\kappa S = 0$.

This concludes the confirmation of fermionic $\kappa$-invariance of our total action in the Green-Schwarz superstring, on our newly developed SYM background in 12D.

## 4. Lorentz Covariant Field Equations

We have so far established an invariant lagrangian under supersymmetry. However, this lagrangian is not 'fully' invariant under Lorentz symmetry in the 12D, due to the usage of null-vectors. We now address our question of Lorentz covariance of our model, by trying to recover the Lorentz symmetry as much as possible, by giving up now the super-invariant lagrangian. The formulation in ref. [9] had neither super-invariant lagrangian nor super-invariant action, but now what we try to present is the set of Lorentz covariant field equations in 12D with supersymmetry.

We first give up the super-invariant lagrangian (2.1), keeping only the result for field equations (2.5) - (2.7). We next try to replace our null-vector $n^\mu$ by something 'covariant'. If we replace it simply by a vector $B_\mu$, then we have to fix its transformation under supersymmetry, as usual in conventional supersymmetric theories. This is rather difficult, because we have to accomplish the closure of supersymmetry, which seems non-trivial. In our 12D, however, there is one way to circumvent this difficulty. Recall the particular form of our 12D algebra of supersymmetry (1.1) with the null-vector $n_\mu$, and also the idea in ref. [14] of identifying $n_\mu$ with a momentum of a second particle. This idea suggests another option of replacing the null-vector with a gradient of some scalar field: $n_\mu \equiv \partial_\mu \varphi$, now without introducing a 'second particle' or non-locality [14][7]. Introduction of such a new field usually poses a problem with closure of supersymmetry. However, this system cleverly avoids this problem, because even if this scalar $\varphi$ does *not* transform under supersymmetry $\delta_Q \varphi = 0$, the closure on $\varphi$ is consistent, due to the relation

$$[\delta_Q(\epsilon_1), \delta_Q(\epsilon_2)]\varphi = [(\bar{\epsilon}_1 \gamma^{\mu\nu} \epsilon_2)(\partial_\nu \varphi)](\partial_\mu \varphi) \equiv 0 \tag{4.1}$$

identically vanishing, thanks to the same gradient $\partial_\mu \varphi$ used both for the translation of $\varphi$ itself and the null-vector!

Note that the scalar field $\varphi$, which is intact under supersymmetry, can still depend on the space-time coordinates in 12D. Recall that we can not use a similar trick in conventional supersymmetric theories. This is because any coordinate-dependent field should transform under supersymmetry, in order to avoid the absurdity that the commutator of two supersymmetry yielding a translation operator should *not* be vanishing, while the commutator vanishes on the field by assumption.

Using this crucial point in mind, we replace all of our null-vectors in our field equations (2.5) - (2.7) with the gradient as $n_\mu \equiv \partial_\mu \varphi$ everywhere in there. We thus get the set of field equations for our field content $(A_\mu{}^I, \lambda^I, C_\mu{}^I, \varphi)$

$$D_\nu \widetilde{F}^{\mu\nu\,I} - f^{IJK}(\bar{\lambda}^J \gamma^{\mu\nu} \lambda^K)(\partial_\nu \varphi) = 0 \;, \tag{4.2}$$

$$(\gamma^{\mu\nu} D_\mu \lambda^I)(\partial_\nu \varphi) = 0 \;, \tag{4.3}$$

$$\widetilde{F}_{\mu\nu}{}^I (\partial^\nu \varphi) = 0 \;, \tag{4.4}$$

where the modified field strength has also the gradient term:

$$\widetilde{F}_{\mu\nu}{}^I \equiv \partial_\mu A_\nu{}^I - \partial_\nu A_\mu{}^I + f^{IJK} A_\mu{}^J A_\nu{}^K - C_\mu{}^I \partial_\nu \varphi + C_\nu{}^I \partial_\mu \varphi \;, \tag{4.5}$$

together with the field equations for $\varphi$:

$$\partial_\mu \partial_\nu \varphi = 0 \;, \tag{4.6}$$

$$(\partial_\mu \varphi)^2 = 0 \;, \tag{4.7}$$

where (4.6) guarantees the constancy of $\partial_\mu \varphi$, while (4.7) guarantees the null-ness of $\partial_\mu \varphi$. A solution for $\varphi(x)$ to these two equations, is fixed to be $\varphi(x) = a + n_\mu x^\mu$ with linear dependence on the extra coordinates $x^{11}$ and $x^{12}$, making the identification $n_\mu \equiv \partial_\mu \varphi$ possible, up to some non-essential overall constant $a$. As has been mentioned, these field equations can be shown to be consistent with the supersymmetry transformation rule

$$\delta_Q A_\mu{}^I = +(\bar{\epsilon} \gamma_\mu \gamma^\nu \lambda^I)(\partial_\nu \varphi) \;, \tag{4.8a}$$

$$\delta_Q \lambda^I = +\tfrac{1}{4} \gamma^{\mu\nu} \epsilon \, \widetilde{F}_{\mu\nu}{}^I \;, \tag{4.8b}$$

$$\delta_Q C_\mu{}^I = -(\bar{\epsilon} \gamma_\mu{}^\nu D_\nu \lambda^I) \;, \tag{4.8c}$$

$$\delta_Q \varphi = 0 \;. \tag{4.8d}$$

Accordingly, the closure of supersymmetry is guaranteed even with (4.8d) or equivalently $\delta_Q(\partial_\mu \varphi) = 0$. All of the equations (4.2) - (4.8) are now manifestly $SO(10,2)$ covariant, as they stand.

We stress that the consistency we have realized in our system is simply due to the particular algebra (1.1), together with the scalar field $\varphi$ intact under supersymmetry. As for the final 'breaking' of Lorentz covariance upon the choice of a solution $\varphi = a + n_\mu x^\mu$, we interpret this as a kind of 'spontaneous breaking' of Lorentz covariance in 12D. In fact, there is no explicit breaking of Lorentz symmetry $SO(10,2)$ for the field equations (4.2) - (4.7), while such breaking is caused only by an explicit non-trivial solution for $\varphi$. In this sense, we can regard this mechanism as 'spontaneous breaking' of the Lorentz symmetry $SO(10,2)$ down to $SO(9,1)$.

Another way to look at the consistency of the solution $\varphi = a + n_\mu x^\mu$ is to consider the $\delta_Q$-variation of this solution. Due to our algebra (1.1), the $\delta_Q$-transformation of the coordinates is $\delta_Q x^\mu = (\bar\epsilon \gamma^{\mu\nu} \theta) n_\nu$ [14], therefore

$$\delta_Q \varphi = \delta_Q(a + n_\mu x^\mu) = n_\mu \left[ (\bar\epsilon \gamma^{\mu\nu} \theta) n_\nu \right] \equiv 0 \tag{4.9}$$

vanishes consistently with (4.8d).

We have also tried to obtain a fully invariant lagrangian that can yield the above set of field equations, but so far we have not reached any consistent lagrangian. For example, just replacing $n_\mu$ by $\partial_\mu \varphi$ in (2.1) does not lead to an totally invariant action. The main problem seems to arise to get the field equation (4.6) with an appropriate lagrange multiplier with a right supersymmetry transformation.[7] At the present time we do not know, if there is to be an invariant lagrangian for the above set of fully covariant field equations.

## 5. Concluding Remarks

In this paper, we have solved the puzzle about the lack of invariant lagrangian for the SYM theory in 12D, which has been lurking since the first establishment of the theory [9]. We have given an explicit lagrangian formulation for the 12D SYM theory for the first time. It has a modified field strength with the Chern-Simons form in terms of the new but auxiliary vector field $C_\mu$ gauging the extra $\Omega$-symmetry. We saw that the multiplet we use do not need any constraint from outside by hand, but is generated automatically as the field equation of the $C_\mu$-field. We found that there are several non-trivial extra symmetries inherent in the system, despite of the simplicity of our lagrangian.

In the original work of $N = 1$ supergravity in 12D [10], addressing the issue of the lack of invariant lagrangian, we took a standpoint that the Green-Schwarz superstring formulation given there will provide an action principle at the level of world-sheet physics. However, the importance of lagrangian formulation within the target space-time is to be stressed, due to the enormous advantage of invariant lagrangian with action principle. As a matter of fact, according to the recent development in M-theory related to matrix theories [15], it seems imperative to establish an lagrangian formulation for these higher-dimensional supergravities in 12D or beyond. This is because such a lagrangian formulation is expected to play a key role to establish the relationship to matrix theories in lower dimensions [15]. In this sense, we imagine the significance of the results given in this paper for any future studies based on higher-dimensional supersymmetry/supergravity theories.

---

[7]Needless to say, the right supersymmetry transformation rule should be closed on-shell.

In the initial work on SYM theory in 12D [9], the origin of the local extra $\Omega$-symmetry was was not clear, but now with the vector field $C_\mu$ gauging this local symmetry, we have more natural construction of the whole theory. We emphasize that this type of Chern-Simons form has never been presented for supersymmetric theories in the past to our knowledge. It may well be that even other extra symmetries such as $\delta_\zeta$ or $\delta_\eta$ for our action have their proper gauge fields, that can also simplify the non-linear transformation structure in the closure of gauge algebra. Even though it should be straightforward, the generalization to higher-dimensional SYM theories [12] might be in practice cumbersome to handle, because of the expected huge set of extra symmetries inherent in the system.

In section 4, we have presented a new manifestly $SO(10,2)$ covariant set of field equations with no explicit use of a null-vector, introducing the new scalar field $\varphi$. Even though this scalar field is intact under supersymmetry, while maintaining its non-trivial coordinate dependence, we have shown how the system avoids the usual problem of closure of supersymmetry. All of our field equations are also manifestly local, with no multi-locality [7][14]. This system realizes what we call 'spontaneous breaking' of Lorentz symmetry, *i.e.*, all the field equations are manifestly covariant both under Lorentz symmetry in 12D and supersymmetry, while the particular choice of solutions breaks the Lorentz covariance. We can call this mechanism also 'on-shell breaking of the Lorentz symmetry $SO(10,2)$', 'spontaneous dimensional reduction', or 'null-vector reduction' which was predicted vaguely in ref. [1]. To our knowledge, there have been no other explicit examples of this sort in the past, and the 12D supersymmetric models and F-theory strongly motivated such a model of Lorentz symmetry breaking. It is interesting that the introduction of a scalar field whose gradient replacing the null-vector is realized consistently by the particular algebra of supersymmetry (1.1) in 12D.

Even though we have not yet succeeded in fixing a fully invariant lagrangian both under supersymmetry and 12D Lorentz transformation, we emphasize that the results above as well as other series of new results [9][10][11][12] are strongly suggestive that there are something deeper underlying our SYM models in generally higher dimensions, which are yet to be explored. For example, even though we did not perform explicitly, application of this technique using scalars for null-vectors to further higher-dimensional SYM theories treated in [12] is also straightforward.

In this paper, we also confirmed is the fermionic $\kappa$-invraiance of the total action for Green-Schwarz superstring, that can provide supporting evidence of the consistency of our 12D SYM backgrounds coupled to superstring. We mention that this $\kappa$-invariance is also consistent with the formulation in section 4 with the introduction of two new scalar superfields $\varphi$, $\widetilde{\varphi}$, whose gradients replace the two independent null-vectors: $n_a \equiv \nabla_a \varphi$, $m_a \equiv \nabla_a \widetilde{\varphi}$.

We stress the difference of our theory from 10D SYM theory, due to the extra components as well as extra coordinate dependence of the Yang-Mills field $A_\mu{}^I$. The constraint $\widetilde{F}_{\mu\nu}{}^I n^\nu = 0$ does *not* delete all the extra components, but there remains the non-vanishing component $\widetilde{F}_{\mu+}{}^I \neq 0$. Even though the derivative $n^\mu D_\mu F_{\rho\sigma}{}^I = D_- F_{\rho\sigma}{}^I = 0$ vanishes due to the Bianchi identity and the constraint $\widetilde{F}_{\mu\nu}{}^I n^\nu = 0$, there is a non-vanishing derivative $D_+ F_{\rho\sigma}{}^I \neq 0$, implying the non-trivial dependence of $\widetilde{F}_{\rho\sigma}{}^I$ on the extra coordinate $X^-$. Additionally, as the inequalities $\Pi_\mp{}^a n_a \neq 0$, $\Pi_\mp{}^a m_a \neq 0$ in our Green-Schwarz formulation indicate, the existence of string variables $X^\pm(\sigma)$ in the extra dimensions depending on one of the world-sheet coordinate $\sigma^\mp$ also provides supporting evidence for the non-triviality of our 12D SYM theory.

In this paper we did not deal with the couplings to supergravity in 12D, but our present result will provide a powerful working ground for a possible fully invariant lagrangian formulation as well as $SO(10,2)$ covariant field equations for $N=1$ supergravity in the future. Studies in this direction as well as other points mentioned above are now under way [20].

We are grateful to I. Bars and C. Vafa for important discussions. Special acknowledgement is due to S.J. Gates, Jr., who stressed the importance of lagrangian formulation.

# References

[1] C. Vafa, Nucl. Phys. **B469** (1996) 403.

[2] I. Bars, Phys. Rev. **D54** (1996) 5203; hep-th/9604200, in Proceedings *'Frontiers in Quantum Field Theory'* (Dec. 1995, Toyonaka, Japan), eds. H. Itoyama *et. al.*, World Scientific (1996), page 52.

[3] D. Kutasov and E. Martinec, Nucl. Phys. **B477** (1996) 652.

[4] I. Bars, Phys. Rev. **D55** (1997) 2373.

[5] I. Bars and C. Kounnas, Phys. Lett. **402B** (1997) 25; *'String & Particle with Two Times'*, hep-th/9705205.

[6] *For reviews, see e.g.,*, M. Duff, *'Supermembranes'*, hep-th/9611203; *'M-Theory (Theory Formerly Known as Strings)'*, hep-th/9608117; J.H. Schwarz, *'Lectures on Superstring and M-Theory Dualities'*, hep-th/9607201.

[7] I. Bars and C. Kounnas, Phys. Rev. Lett. **77** (1996) 428; *'A New Supersymmetry'*, hep-th/9612119.

[8] I. Bars, Phys. Lett. **403B** (1997) 257.

[9] H. Nishino and E. Sezgin, Phys. Lett. **388B** (1996) 569.

[10] H. Nishino, *'Supergravity in 10+2 Dimensions as Consistent Background for Superstring'*, hep-th/9703214.

[11] H. Nishino, *'N = 2 Chiral Supergravity in (10+2)-Dimensions as Consistent Background for Super (2+2)-Brane'*, UMDEPP 97-122, hep-th/9706148.

[12] H. Nishino, *'Supersymmetric Yang-Mills Theories in $D \geq 12$'*, UMDEPP 98-006, hep-th/9708064.

[13] E. Sezgin, Phys. Lett. **403B** (1997) 265.

[14] I. Bars and C. Deliduman, USC-97/HEP-B5, CERN-TH/97-181, hep-th/9707215; USC-97/HEP-B6, hep-th/9710066.

[15] T. Banks, W. Fischler, S.H. Shenker and L. Susskind, Phys. Rev. **D55** (1997) 5112; V. Periwal, Phys. Rev. **D55** (1997) 1711.

[16] H. Nishino and E. Sezgin, Phys. Lett. **144B** (1984) 187; Nucl. Phys. **B144** (1986) 353; hep-th/9703075, to appear in Nucl. Phys. B.

[17] H. Nishino and S.J. Gates, Jr., Int. Jour. Mod. Phys. **A8** (1993) 3371.

[18] S.J. Gates, Jr., M.T. Grisaru, M. Roček and W. Siegel, *'Superspace'* (Benjamin/Cummings, 1983).

[19] J.J. Atick, A. Dhar and B. Ratra, Phys. Lett. **169B** (1986) 54; E. Bergshoeff, E. Sezgin and P. Townsend, Phys. Lett. **169B** (1986) 191.

[20] H. Nishino, *'Supergravity Theories in $D \geq 12$'*, UMD preprint, in preparation.